\documentclass[runningheads]{svmult}

\usepackage{makeidx}   
\usepackage{graphicx}  
\usepackage{subeqnar}  
\usepackage{multicol}  
\usepackage{physprbb}  
\makeindex             

\newcommand{\farcs}{\hbox{$.\!\!^{\prime\prime}$}}

\begin{document}
\title*{Measuring the mass of high-$z$ galaxies with NGST}
\toctitle{Measuring the mass of high-$z$ galaxies with NGST}
%
%
\titlerunning{high-$z$ galaxies with NGST}
%
\author{ Tommaso Treu\inst{1}
\and Massimo Stiavelli\inst{2}}
\authorrunning{Treu \& Stiavelli}
%
%
\institute{
California Institute of Technology, Astronomy 105-24, Pasadena CA 91125
\and 
Space Telescope Science Institute, 3700 San Martin Dr,  Baltimore MD 21218}
\maketitle              

\begin{abstract}
We discuss dynamical mass measurements of high-$z$ galaxies with the
Next Generation Space Telescope (NGST). In particular, we review some
of the observational limits with the current instrument/telescope
generation, we discuss the redshift limits and caveats for absorption
and emission lines studies with NGST, and the existence of suitable
targets at high redshift. We also briefly summarize strengths and
weaknesses of proposed NGST instruments for dynamical studies.
\end{abstract}

\section{Introduction}

During this meeting we have heard of mass measurements at redshifts
that were out of the realm of possibilities just a few years
ago. However, several obstacles stand between us and the future when
dynamical mass measurements will be routinely feasible at $z>1$. Among
them, the lack of spatial resolution and the atmospheric
emission/absorption at infrared (IR) wavelengths (see e.g. the
contribution by M.~Franx). The Next Generation Space Telescope (NGST)
-- with its unique combination of large collecting area, superb
spatial resolution, and low background -- will provide a major
contribution to the extension of mass measurements at $z>1$.

Here we discuss mass measurements with NGST, focusing on dynamical
measurements (lensing and stellar mass measurements are discussed
elsewhere in these proceedings, see e.g. the contribution by
H.~Ferguson).  As for any dynamical mass measurement, a spatial scale
and a velocity scale are needed. In the following, we will assume that
spatial scales are easily measurable even though this is not the case
for sub-galactic clumps at very high redshift. Our focus in this
contribution will be on emission and absorption line measurements
(Sec.~2 and~3) of velocity scales.  In particular, we will briefly
review the limits of what is feasible with the current technology and
present some detailed simulations on the capabilities of NGST for mass
measurements. Finally, we will discuss the implications of the choice
of NGST-instrumentation for mass measurements.

\begin{figure}[b]
\begin{center}
\includegraphics[width=0.4\textwidth]{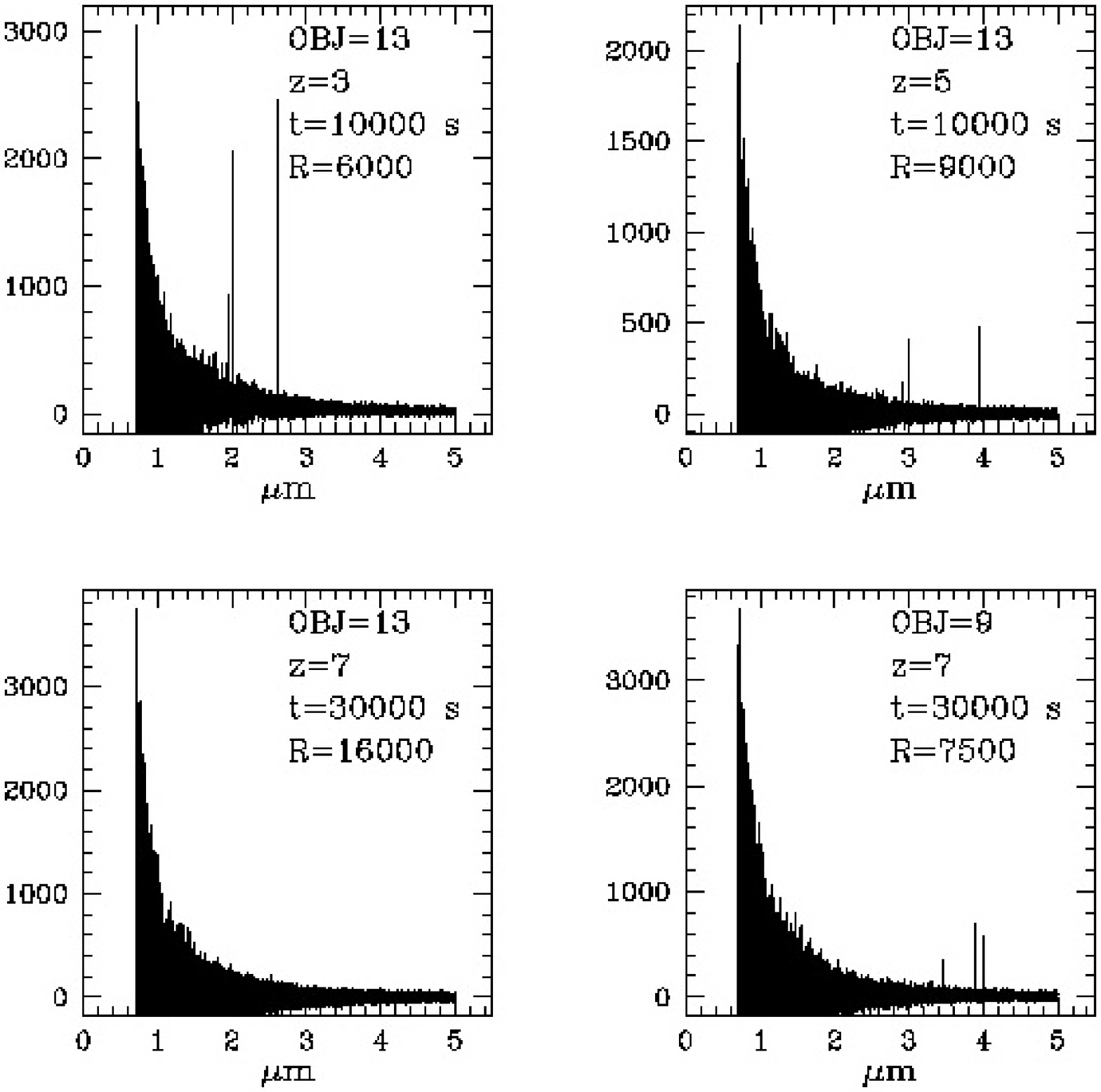}
\includegraphics[width=0.43\textwidth]{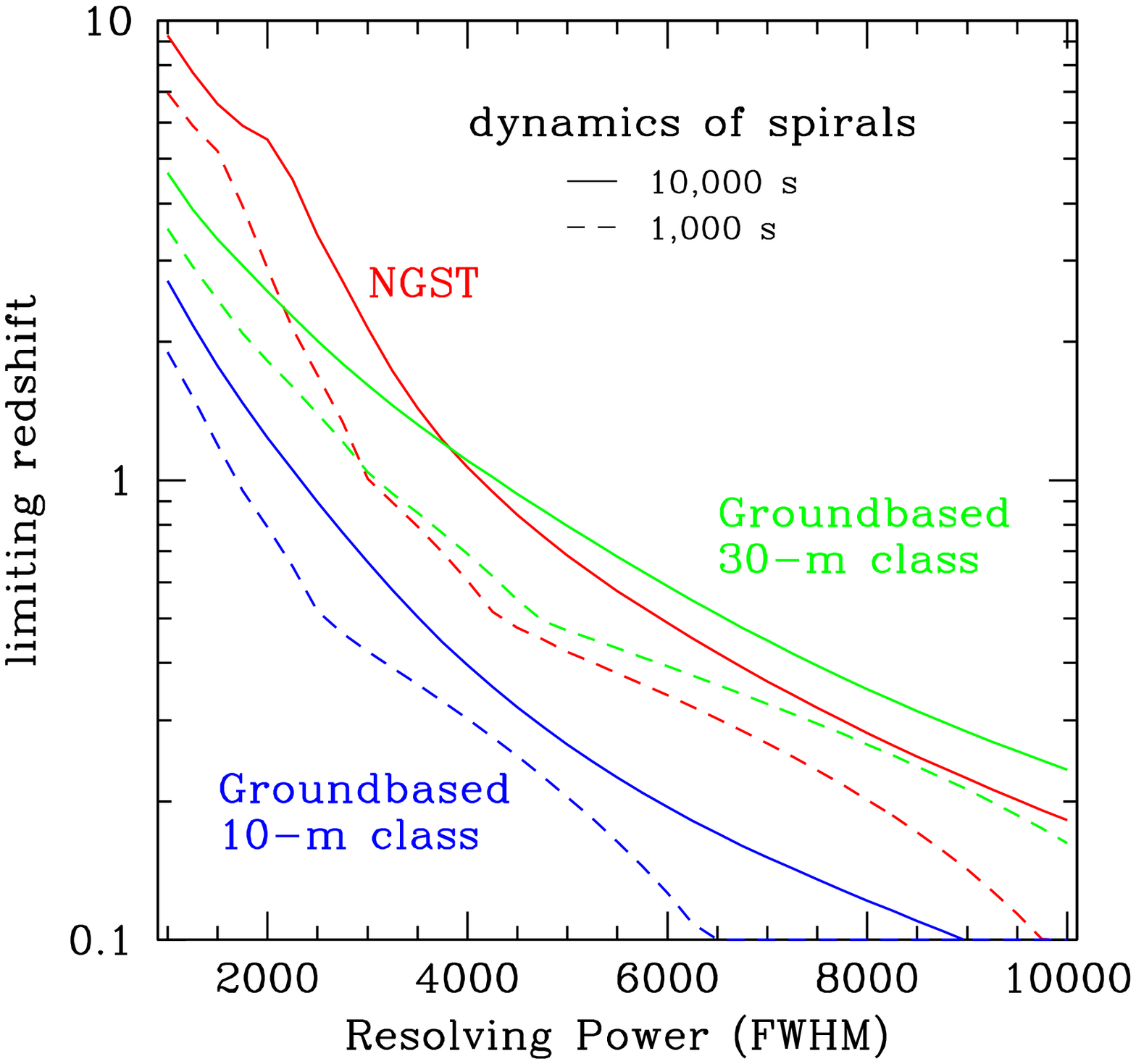}
\end{center}
\caption[]{Left panels: simulated spectra for the typical (OBJ=13)
Milky Way progenitor and for the brightest (OBJ=9) Milky Way
progenitor. Resolving powers in excess of 6,000 are needed for z$>$3
even for the most massive progenitors of L$\star$ galaxies. Right
panel: correlation between available resolving power and maximum
redshift for which a the internal kinematics of a spiral galaxy can be
resolved. For relatively large objects at high redshift NGST is
superior to even a 30m ground based telescope. }
\label{fig:sim}
\end{figure}

\section{Emission Line Measurements}

Kinematical measurements based on emission lines are easy to carry out
but hard to interpret. This is because the gas may be far from
equilibrium and thus gas kinematics may not be telling us anything
about the mass of the host galaxy. One way to overcome this difficulty
is to have access to two-dimensional kinematical information and this
is indeed what was needed to solve the same problem in the context of
mass measurement in nearby spiral galaxies using HI \cite{vanAlbSan}
and black hole mass measurements from the kinematics of nuclear gas
disks \cite{Harms}. NGST presents several advantages in this context
because of its high angular resolution. Its uninterrupted wavelength
coverage guarantees that strong emission lines will be available over
a large range of redshifts. The low background redwards of 2.5 $\mu$m
will guarantee high sensitivity for measurements using H$\alpha$ at
z$\geq$ 3. On the basis of the visibility of H$\alpha$ one can argue
that for objects at z$\leq$ 3 a detailed 2D kinematic mapping can be
carried out from the ground using an adaptive optics fed integral
field spectrograph on an 8-meter class telescope. At z$\geq$3 oxygen
lines are still accessible from the ground but may be suppressed by
the expected low metallicity of most objects at that redshift. Thus,
the availability H$\alpha$ and the sensitivity of NGST make it the
ideal instrument at z$\geq$3.

In addition to the availability of lines another issue is the
resolving power needed to carry out a measurement. In a nutshell,
faint galaxies will on average have lower mass and smaller internal
velocities, requiring higher resolving power to be studied. Since on
average we expect galaxy mass to decrease with redshift, one will need
progressively higher resolving power to study galaxies at increasingly
high redshift. This intuitive result is illustrated in
Figure~\ref{fig:sim} where we show, in the left panel, simulated NGST
spectra for a typical Milky Way progenitor (OBJ=13) at z=3, 5, and 7
and for a bright Milky Way progenitor (OBJ=9) at z=7. These models
have been obtained with a merging tree code \cite{withMike,Liege}. The
resolving power needed to measure internal kinematics is in the range
6,000 to 16,000. In the right panel we show a relation connecting the
available resolving power to the highest redshift that can be
probed. This plot has been obtained by assuming the validity of the
Tully-Fisher relation and by requiring that for the given exposure
time and resolving power sufficient signal-to-noise is achieved to
carry out the measurement. The optimal emission lines is used for each
redshift \cite{Goddard}. It is clear that for low mass objects NGST is
not competitive with large ground based telescopes. However, NGST is
superior to 30m class ground based telescopes for massive objects at
$z>2$.

\begin{figure}[h!]
\begin{center}
\includegraphics[width=\textwidth]{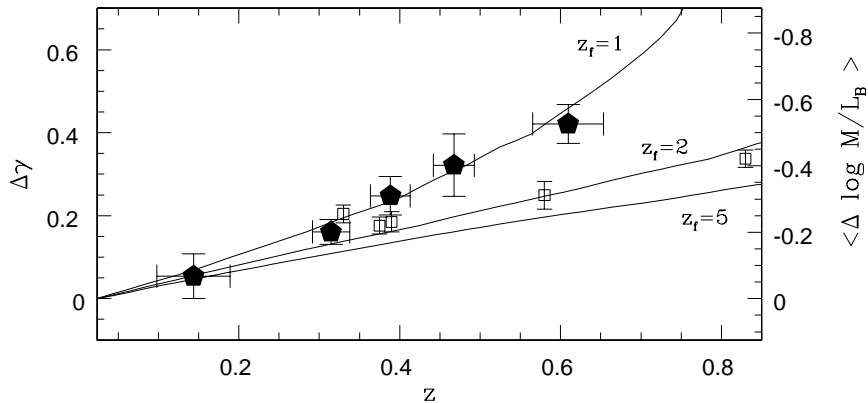}
\end{center}
\caption[]{Absorption line dynamics at $0.1<z<1$, ground-based limits:
evolution of the Fundamental Plane of E/S0 galaxies from $z=0.8$ to
$z=0$ (from \cite{T02}).  The average offset of the intercept of field
galaxies from the local FP relation as a function of redshift (large
filled pentagons), is compared to the offset observed in clusters
(open squares). See \cite{T02} for references, description, and
details.}
\label{fig:FP}
\end{figure}

\section{Absorption Line Measurements}

Stellar absorption lines are a very good probe of velocity scales for
several reasons: {\it i)} a large fraction of galaxies do not have
emission lines, and therefore absorption lines are the only way to
go; {\it ii)}  stars are a good tracer of the velocity distribution of
the system, as opposed to emission lines coming from HII regions which
might not be in dynamical equilibrium with the galaxy; {\it iii)} stellar
absorption lines kinematics tend to suffer less from the effects of
interstellar absorption, local motions, and winds then emission lines
kinematics; {\it iv)} if streaming motions are not significant as in
massive E/S0 galaxies, interesting dynamical constraints can be
gathered without the need for spatially resolved information.

However, stellar absorption lines have to be present, and stellar
populations need to be in dynamical equilibrium, in order to obtain
meaningful information. Typical optical absorption lines (e.~g. the Mg
triplet) need of order $\sim 1$~Gyr to develop in stellar populations,
while the time scales can be relatively shorter for other frequently
used lines such as the near-IR Ca triplet. On the one hand this is
good, because it guarantees that if such lines are present then the
stellar populations are almost certainly old enough that the system is
in equilibrium. On the other hand, it is worth asking the question of
whether systems with stellar absorption lines exist at high redshift.
In the following we will illustrate with a few examples the current
limits of ground based measurements, we will show why we believe there
are interesting targets beyond these limits, and we will explore the
feasibility of such measurements with NGST.

\begin{figure}[b]
\begin{center}
\includegraphics[width=0.4\textwidth]{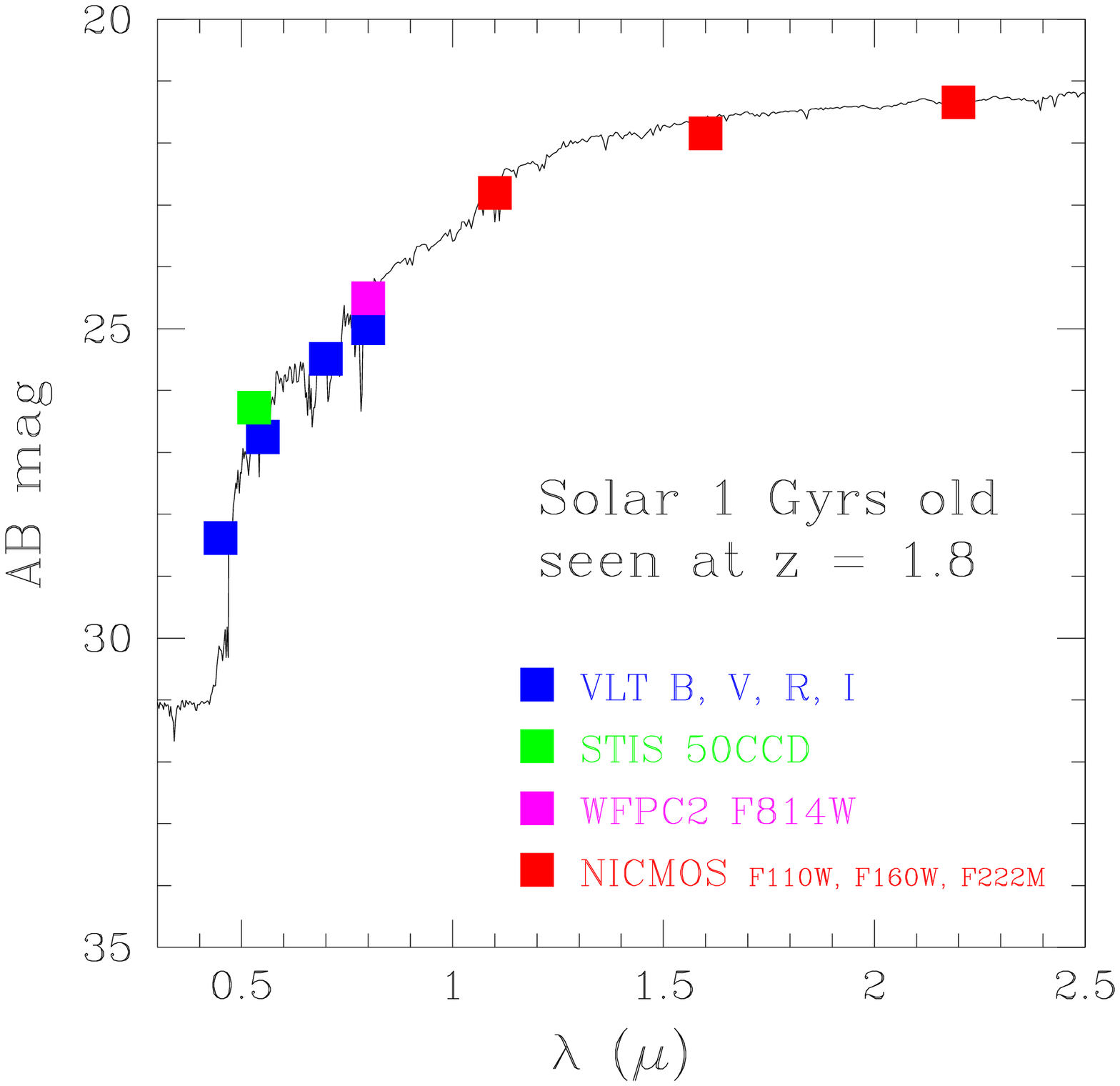}
\includegraphics[width=0.43\textwidth]{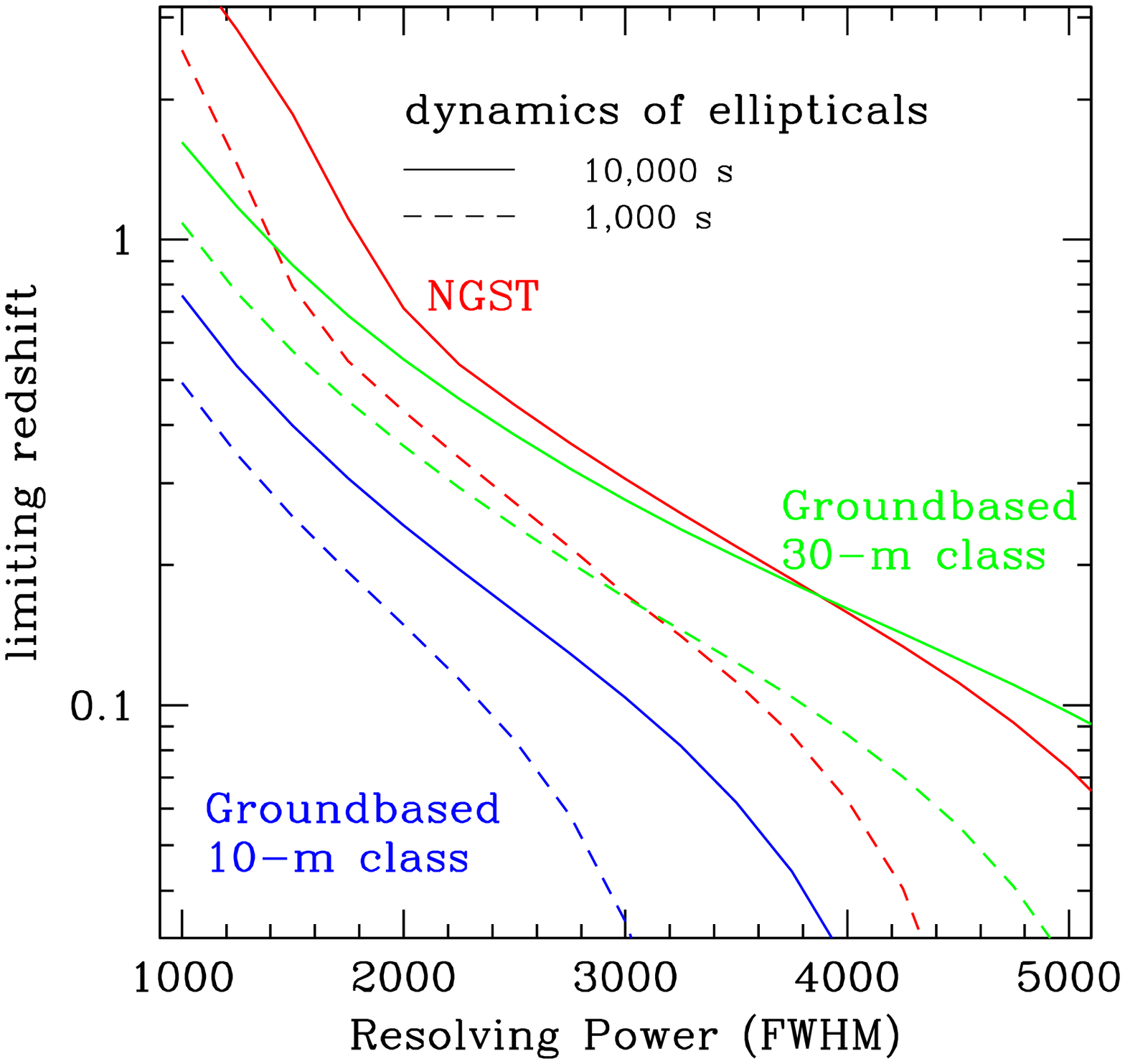}
\end{center}
\caption[]{Absorption lines dynamics at $z>1$: targets for NGST. Left:
spectral energy distribution of and extremely red object with E/S0
morphology at $z\sim1.8$ (from \cite{T98,S99}). Right: comparison of
the performance of NGST, a 10m ground based telescope and a 30m ground
based telescope.}
\label{fig:highzES0}
\end{figure}

Recent studies of the evolution of the Fundamental Plane (FP) with
redshift show that E/S0 galaxies do not undergo major structural
changes from $z\sim0.8$ to the present
(e.~g. \cite{vd98,T01b,T02}). Such measurements, based on a
combination of spectroscopy (velocity dispersion) and imaging
(photometric structural parameters) provide important information on
the evolution of the internal structure of E/S0 galaxies and their
stellar populations (see also the contribution by G.~Illingworth). For
example, \cite{vd98} inferred from the evolution of the FP that
cluster E/S0 have old stellar populations, while \cite{T02} used this
technique to show that secondary episodes of star formation are common
in massive field E/S0 at $z\sim0.5$. It would be interesting to extend
such measurements to higher redshifts, where we know that some old
E/S0 exist, at least in the range $z=1-2$ (\cite{K20,T98,S99}; see
Figure~\ref{fig:highzES0}).  However, long integrations on large
telescopes are needed to push such studies significantly beyond
$z\sim1$ because E/S0 become very faint in the optical and sky
emission lines severely limit what can be done in the IR. For example,
\cite{KT02} measured the central velocity dispersion of the lens
galaxy in the gravitational lens system MG2016+112 ($z=1.004$) by
integrating 8.5hrs with the Echellette Spectrograph Imager (ESI) at
the Keck II telescope. Given the excellent sensitivity and resolution
of ESI (that allows for a good removal of sky lines in the red/IR), it
seems likely that with the current generation of 8-10m class
telescopes such measurements will be limited to bright E/S0 at
$1<z<1.5$.

As shown in the right panel of Fig.~\ref{fig:highzES0}, the redshift
range $z=1-2$ would be easily within reach with NGST (see also
\cite{T98a}), and even objects beyond $z\sim2$ could be targeted, if
they exist. Based on calculations similar to the one used in Fig.~1
(right panel), adopting the FP instead of the Tully-Fisher relation as
luminosity-velocity scaling law, at $z>1$ NGST will be significantly
better than a 30-m telescope, especially for large masses.  Recent
studies suggest that old stellar populations indeed exist at
$z>2$. For example \cite{S01} found that the colors of Lyman-$\alpha$
selected galaxies at $z\sim2.4$ are consistent with those of an old
stellar population ($\sim$ Gyr of age) -- responsible for the observed
red colors -- with a sprinkle of starbursting population --
responsible for the observed emission lines -- (see also \cite{F01}).

\subsection{The Internal Mass Distribution of High Redshift Galaxies}

So far, we only discussed absorption line kinematics in the absence of
spatially resolved information. However, spatially resolved kinematics
have been very important in our study of the local Universe.  For
example it has provided crucial constraints on the internal structure
of E/S0 (e.~g. \cite{B94,G01}), and Hubble Space Telescope (HST)
subarcsecond resolution spectroscopy has been used to prove the
existence of black holes at the center of massive E/S0. Unfortunately,
the small aperture of HST and the seeing-limited spatial resolution of
ground based telescopes have so far basically confined spatially
resolved kinematics to the local Universe. As an example of the
current observational limits, \cite{TK02} have recently started a
kinematic survey of gravitational lenses. Under the best observing
conditions (seeing $0\farcs6$) -- with long integration times at the
Keck-II Telescope -- they were able to measure velocity dispersion
profiles extended beyond the effective radius for lens galaxies at
$z<0.5$. Although the extended profiles provide valuable and unique
information on the mass distribution, a higher spatial resolution is
needed to study in detail the mass distribution within the central
kiloparsecs. NGST -- with its combination of large collecting area and
superb spatial resolution -- offers the tremendous opportunity to
explore spatially resolved kinematics at cosmological distances. With
a typical resolution $\leq0\farcs1$ NGST will be able to study in
detail the mass distribution of E/S0 and its evolution with redshift.

\section{Mass Measurements with NGST: Instrumentation}

Both emission and absorption line studies of the kinematics of high
redshift galaxies require a resolving power of at least 3,000,
i.e. somewhat higher than what is currently in the baseline for
NGST. The need for high resolving power is stronger for emission line
measurements since they can be carried out on intrinsically fainter
objects. Because of this reason there is a strong rationale for a
two-dimensional spectroscopic capability at R$\simeq$3,000. This
capability could be in the form of an integral field spectrograph
\cite{Bacon} if budget, volume, and weight constraints allow it.
Alternatively, it could be provided by a multi-object spectrograph
based on a configurable slit array provided the configuration time is
short enough. Such a spectrograph would allow one to step the slit
over resolved objects to be mapped in two dimensions while integrating
on fainter sources \cite{3D}.

\section{Acknowledgments}

We acknowledge financial support by NASA through a grant from STScI
(HST-AR-09222), which is operated by AURA, under NASA contract
NAS5-26555.

%

\end{document}